\documentstyle[12pt]{article}
\newfont{\myfont}{cmr12 scaled \magstep1}
\begin{document}
\baselineskip=25pt
\title{\bf
Spin  Gap of Two-Dimensional Antiferromagnet Representing CaV$_4$O$_9$ }
%\bigskip
\author{ Kazuhiro Sano ~and~ Ken'ichi Takano$^1$ \\
 {\it Department of Physics, Faculty of Education, Mie University,} \\
  {\it Tsu, Mie 514 }     \\
$^1${\it Toyota Technological Institute, Tenpaku-ku, Nagoya 468}}
\date{(Received \hspace{5cm})}

\maketitle

     We examined a two-dimensional Heisenberg model with two kinds of
exchange energies, $J_e$ and $J_c$.
     This model describes localized spins at vanadium ions in a layer of
CaV$_4$O$_9$, for which a spin gap is found by a recent experiment.
     Comparing the high temperature expansion of the magnetic
susceptibility to experimental data, we determined the exchange energies as
$J_e \simeq$ 610 K and $J_c \simeq$ 150 K.
     By the numerical diagonalization we estimated the spin gap as $\Delta
\sim 0.2J_e \simeq$ 120 K, which consists with the experimental value 107
K.
     Frustration by finite $J_c$ enhances the spin gap.


\newpage

     Low-dimensional antiferromagnets with spin gap attract much interest
due to the possible relevance to the high-$T_c$ superconductivity.
     For one dimension, many systems with spin gap have been examined
theoretically.
     For example, the ladder model opens a spin gap up to about
0.5$J$.~\cite{Dagotto,Gopalan,Troyer}
     The Majumdar-Ghosh model, a typical frustrated system, has a spin gap
of about
0.24$J$.~\cite{MGmodel1,MGmodel2,Shastry,Tonegawa,Sano}
     A series of Heisenberg models with linearly decreasing exchange
interactions also have finite spin gaps in the range between 0.25$J$ and
0.75$J$.~\cite{KTakano1,KTakano2}
     In the above, $J$ is a typical exchange energy included in each model.
     Experimentally, finite spin gaps are found in (VO)$_2$P$_2$O$_7$
\cite{Barnes} and SrCu$_{2}$O$_{3}$.~\cite{MTakano,Ishida}
     These materials are represented well as ladder spin systems arranged
in parallel and coupled weakly.

     For two dimensions, Taniguchi et al. \cite{Taniguchi} recently found a
finite spin gap for a layered material CaV$_4$O$_9$.
     They estimated the spin gap as $\Delta \simeq$ 107 K by measuring the
magnetic susceptibility of $d$-electron spins at vanadium ions (V-spins)
and the spin-lattice relaxation rate of $^{51}$V nuclear moment.
     As long as we know, this is the first experiment showing clearly
the spin gap for (quasi-) two-dimensional spin systems.
     The lattice structure of CaV$_4$O$_9$ shows that there are two kinds
of important exchange interactions between edge sharing V-spins and between
corner sharing V-spins in a layer.
     We denote the corresponding exchange energies as $J_e$ and $J_c$,
respectively.
     Katoh and Imada \cite{Katoh} examined the spin gap by assuming $J_c$ = 0.
     They estimated as $\Delta \simeq 0.11J_e$ by the Quantum Monte Calro
simulation and explained the ground state character by a perturbation
calculation, although they did not estimate effects of $J_c$.
     Ueda et. al.~\cite{Ueda} carried out a similar perturbation
calculation for finite $J_c$.
     However, since these perturbation calculations are not justified for
realistic parameter values, their results have not been definite yet.

     In this letter, we construct a reasonable spin model with realistic
values for $J_e$ and $J_c$ and explain the measured spin gap for V-spins in
CaV$_4$O$_9$.
     The exchange energies are determined by carrying out the high
temperature expansion (HTE) of the magnetic susceptibility for the model
and by comparing them to experimental data.
     We examine the spin gap by numerically diagonalizing the Hamiltonian
for finite systems.
     After the extrapolation to an infinite system is taken, the estimated
spin gap is shown to be fairly close to the experimentally obtained spin
gap.

     To construct the Hamiltonian, we survey the structure of a layer in
CaV$_4$O$_9$.~\cite{Taniguchi,Bouloux}
     A layer consists of VO$_5$ pyramid-shaped clusters with apical oxygen
atoms above and below a basal plane.
     Each pyramid contains a vanadium ion V$^{4+}$ roughly in its center.
     An electron in the $d\epsilon_{xy}$ orbital of a V$^{4+}$ ion forms a
localized spin (V-spin).~\cite{d-orbital}
     Superexchange interactions may occur between V-spins in edge sharing
pyramids and between V-spins in corner sharing pyramids by hybridization of
$d\epsilon_{xy}$ orbitals with $p_{x}$ or $p_{y}$ orbitals of adjacent
oxygens.
     These  superexchange interactions contribute to $J_e$ and $J_c$
respectively.
     There may be also a direct exchange interaction between V-spins in
edge sharing pyramids  due to the overlap of the $d\epsilon_{xy}$ orbitals,
which contributes to $J_e$.
     Since it is difficult to calculate the values of $J_e$ and $J_c$ by
starting from the first principle, we determine them by comparing
experimental data of the magnetic susceptibility to the HTE calculation, as
will be shown.
     Thus we describe magnetic properties of CaV$_4$O$_9$ by a
two-dimensional antiferromagnetic Heisenberg model with two kinds of
exchange energies $J_e$ and $J_c$.

     The Hamiltonian for CaV$_4$O$_9$ is written as
\begin{equation}
     H=H_0+H_A+H_B ,
\end{equation}
\begin{equation}
   H_0 = J_e \sum_{<i,\alpha>}{\bf S}_{i}\cdot {\bf S}_{\alpha} , \quad
   H_A = J_c \sum_{<i,j>}{\bf S}_{i}\cdot {\bf S}_{j} , \quad
 H_B = J_c \sum_{<\alpha,\beta>}{\bf S}_{\alpha}\cdot {\bf S}_{\beta} ,
\end{equation}
where  ${\bf S}_{i}$ (${\bf S}_{\alpha}$) is the V-spin at site $i$
($\alpha$) belonging to the $A$ ($B$) sublattice.
     The lattice structure in a layer is shown in Fig. 1(a).
     A small circle (square) represents V-spins above (below) the basal plane.
     We have called the set of sites denoted by small circles (squares) the
$A$ ($B$) sublattice.
      The exchange interactions for $\langle i,\alpha \rangle$ is denoted
by bold solid line in Fig. 1 (a);  the corresponding exchange energy is
$J_c$.
      The exchange interactions for $\langle i,j \rangle$ ($\langle
\alpha,\beta \rangle$) are denoted by thin (dashed) solid lines;  the
exchange energy is $J_e$.
     We note that the sub-Hamiltonian $H_A$ ($H_B$) consists only of spins
belonging to the $A$ ($B$) sublattice, while $H_0$ consists of spins
belonging to both.
     In the case of $J_c=0$, the Hamiltonian reduces simply to $H_0$.
     The lattice for $H_0$ given by thin solid lines in Fig. 1(a) is
topologically the same as the lattice of Fig. 2.
     In the case of $J_e=0$, spins on the A sublattice are described only
by the sub-Hamiltonian $H_A$ and do not interact with spins on the B
sublattice, which are described only by $H_B$.
     Hence we can consider $H_A$ and $H_B$ separately.
     The sublattice for $H_A$ ($H_B$) represented by thin solid (dashed)
lines in Fig. 1(a) is topologically the same as the lattice shown in Fig.
2.
     Thus all sub-Hamiltonians $H_0$, $H_A$ and $H_B$ are equivalently
represented by the unfrustrated lattice of Fig. 2.
     The ratio $\gamma = J_c/J_e$ changes the strength of frustration;
     the system is unfrustrated both in the limits of $\gamma = 0$ and
$\gamma = \infty$.

     To determine the exchange energies, $J_e$ and $J_c$, we carry out the
HTE for magnetic susceptibility and compare the result to experimental
data.~\cite{Taniguchi}
     The experimental susceptibility $\chi_E$ is shown in Fig. 3 as a
function of $1/T$.
     The HTE is known to precisely reproduce a high-temperature part of a
thermodynamic quantity of quantum spin systems.~\cite{Rushbrooke}
     We derived the HTE formula for the magnetic susceptibility $ \chi_{H}$
of the Hamiltonian (1).
It is written as
$$
     \chi_{H}=C\biggl[\frac{1}{T}-\frac{3(1+\gamma)J_e}{4T^2}
     + \frac{3(1+6\gamma+\gamma^2)J_e^2}{16T^3}\biggr]
     + O\biggl(\biggl(\frac{1}{T}\biggr)^4 \biggr) ,
\eqno{(3)}
$$
where $C = n (g \mu_B)^2/4k$ with $n$ being the number of vanadium ions per
gram in CaV$_4$O$_9$, $g$ the $g$-value, $\mu_{\rm B}$ the Bohr magneton
and $k$ the Boltzmann constant.
    To precisely compare the HTE susceptibility (3) to the experimental
data, we introduce the quantity,
${\tilde{\chi}}_H=(\chi_{H}-\frac{C}{T})T^2$.
    Then eq. (3) gives a linear function of $1/T$ for ${\tilde{\chi}}_H$ as
${\tilde{\chi}}_H= c_0 + c_1/T$ with $c_0 = - \frac{3}{4} (1+\gamma)J_e$
and $c_1 = \frac{3}{16} (1+6\gamma+\gamma^2)J_e^2$.
     We plotted the corresponding experimental quantities
$\tilde{\chi}_{E}=(\chi_{E}-\frac{C}{T})T^2$ as a function of $1/T$ in the
inset of Fig. 3.
     The coefficients, $c_0$ and $c_1$, are determined so that
${\tilde{\chi}}_H$ represents the tangential line of $\tilde{\chi}_{E}$, as
shown in the inset;  i.e. $c_0$ = $-$2.21 emu/gK$^2$ and $c_1$ = 693
emu/gK.
     From the coefficients, we obtain $J_e \simeq 610$ K, $J_c \simeq 150$
K and then $\gamma \simeq 0.25$.~\cite{superexchange}
     Using these values we plotted the $\chi_{H}$ in Fig. 3 by the dashed line.
     The exchange energy of $J_e \simeq 610$ K seems to be fairy large in
vanadium oxides and  becomes a half  of cuprates.~\cite{Ishida}
     The value of $J_c$ is smaller than $J_e$ but is not negligible, so
that the real system is frustrated.
     We should carefully consider the contribution of $J_c$ when we examine
the magnetic properties of CaV$_4$O$_9$.

      To obtain the spin gap, we numerically diagonalize the Hamiltonian
(1)  by Lanczos' method for finite systems with the periodic and/or the
antiperiodic boundary conditions.
      In the case of $\gamma = 0$ we can use the systems with $N$= 12, 16,
18 and 24, which are shown in Fig. 2.
      However in the case of finite $\gamma$, systems with $N$ = 16 and 24
among them only fit to the boundary conditions;  these finite systems are
shown in Fig. 1(b).
     This is because the lattice of $\gamma = 0$ (Fig. 2) is more symmetric
than that of $\gamma \neq 0$ (Fig. 1(a)).
     We calculate the excitation energies $\Delta$ from the singlet ground
states to the lowest triplet states both for $\gamma = 0$ and $\gamma =
0.25$ and compare them.
     For the extrapolation, we assume the system-size dependence of $
\Delta \sim 1/N $.
     The results are shown in Fig. 4.
     Data for $\gamma = 0$ fit a straight line well and confirms the
system-size dependence.
     Hence the spin gap for $\gamma = 0$ is estimated as $\Delta \simeq
0.13J_e$ in the thermodynamic limit.
     This agrees with the result obtained by the Quantum Monte Carlo
calculation.~\cite{Katoh}
     We expect this system-size dependence is correct even for small but
finite $\gamma$ and apply it to the realistic case of $\gamma = 0.25$.
     Then the spin gap for $\gamma = 0.25$ is estimated as $\Delta \sim
0.2J_e$ in the thermodynamic limit, as shown in Fig. 4.
     This result shows that frustration remarkably enhances the spin gap.
     Using the exchange energy $J_e \simeq 610$ K, the spin gap is
evaluated as $\Delta \sim$ 120 K.
     This result is fairly close to the experimentally obtained spin gap 107 K.

     We finally discuss the origin of the spin gap.
     First we consider the case of $J_c=0$ represented by the lattice of Fig.
2.
     The Hamiltonian $H_0$ consists of two different types of interaction
bonds.
     An interaction bond is one of four bonds forming a plaquette and
another connects two plaquettes.
     We call them a plaquette bond and a dimer bond, respectively.
     Accordingly $H_0$ is written as  $H_0 = H_{0p} + H_{0d}$, where
$H_{0p}$ ($H_{0d}$) consists only of plaquette (dimer) bonds.
     We calculated the expectation values of $H_{0p}$ and $H_{0d}$ in the
ground state by the numerical diagonalization.
     In the thermodynamic limit these values are $\langle H_{0p} \rangle
\simeq -0.20J_eN$ and $\langle H_{0d} \rangle \simeq -0.07J_eN$, when the
numerical results are extrapolated as a function of $1/N$.
     Hence the portions of the energy gains in plaquette parts and in dimer
parts are ${\langle H_{0p} \rangle}/{\langle H_{0} \rangle} \simeq 0.74$
and ${\langle H_{0d} \rangle}/{\langle H_{0} \rangle} \simeq 0.26$,
respectively.
     This result agrees with the picture in which singlet clusters are
formed in plaquette parts.~\cite{Katoh,Ueda}
     Next we consider the general case of $J_c \neq 0$ in terms of the
spin-spin correlation function $\langle {\bf S}_i \cdot {\bf S}_j \rangle$.
     This quantity is evaluated by the numerical diagonalization and the
extrapolation, again.
     At a plaquette bond we obtained $\langle {\bf S}_i \cdot {\bf S}_j
\rangle \simeq -$0.60 for $\gamma=0.25$ against $\langle {\bf S}_i \cdot
{\bf S}_j \rangle \simeq -$0.54 for $\gamma=0$.
     At a dimer bond we obtained $\langle {\bf S}_i \cdot {\bf S}_j \rangle
\simeq -$0.24 for $\gamma=0.25$ against $\langle {\bf S}_i \cdot {\bf S}_j
\rangle \simeq -$0.39 for $\gamma=0$.
     This result shows that parts gaining the correlation energy move from
dimer bonds to plaquette bonds as $\gamma$ increases.
     This tendency corresponds to the enhancement of the spin gap with
frustration.

%\section*{Acknowledgement}
      We would like to thank Satoshi Taniguchi and Masatosi Sato for useful
discussion and presentation of detailed experimental data.
      We also acknowledge Takami Tohyama for discussion on the electronic
structure of CaV$_4$O$_9$.
      One of us (K. T.) carried out this work partially in Department of
Physics of Nagoya University as a Guest Associate Professor.
\clearpage
%\newpage

%
\clearpage

\noindent
{\large \bf Figure Captions}

\begin{description}

\item[Fig. 1.]
     (a) The lattice structure for V-spins in a layer of CaV$_4$O$_9$.
     Small circles (squares) represent V-spins belonging to the $A$ ($B$)
sublattice and are in pyramids above (below) the basal plane.
     The exchange energy for a bold solid line is $J_e$.
     Those for a thin solid line and a thin dashed line are $J_c$.
     (b) Parts of the lattice used in the numerical diagonalization.

\medskip
\item[Fig. 2.]
     The lattice representing a sub-Hamiltonian $H_{0}$, which becomes the
total Hamiltonian when $J_c =$ 0.
     This lattice also represents each of $H_A$ and $H_B$, which are
separated from each other when $J_e =$ 0.
     Parts of the lattice used in the numerical diagonalization are also
shown by the thin solid line.

\medskip
\item[Fig. 3.]
      Magnetic susceptibilities of the experiment\cite{Taniguchi} and the
HTE calculation.
     The HTE curve is determined so that ${\tilde{\chi}}_H$ becomes a
tangential line of ${\tilde{\chi}}_E$ as shown in the inset.

\medskip
\item[Fig. 4.]
     System size dependence of the spin gap $\Delta$ for $\gamma$ = 0
(circles) and 0.25 (squares).
     The extrapolations are done with the solid and the dashed lines
determined by the  method of least  squares with data of $N \geq 16$.

\end{description}


\begin{thebibliography}{99}
\bibitem{d-orbital}
     We did not take account of $d\epsilon_{xz}$ and $d\epsilon_{yz}$
orbitals, since any anomaly indicating the Jahn-Teller effect  is  not
observed in the magnetic susceptibility for 5 K $\displaystyle{
\mathop{<}_{\sim}}$ $T$ $\displaystyle{ \mathop{<}_{\sim}}$ 700 K.
\end{thebibliography}
\end{document}